# Reconstruction of optical parameters for molecular tomographic imaging

Wenxiang Cong, Xavier Intes, Ge Wang

Biomedical Imaging Center, Department of Biomedical Engineering

Rensselaer Polytechnic Institute, Troy, New York 12180

**Abstract:** Optical molecular tomographic imaging is to reconstruct the concentration distribution of photon-molecular probes in a small animal from measured photon fluence rates. The localization and quantification of molecular probes is related to tissue optical properties. The optical parameters can be reconstructed by the optimization method, in which the objective function is not convex, and there are multiple local optimal solutions. The gradient-based optimization algorithm could converge prematurely to an inaccurate solution if the initial guess is not close enough to the true solution. In this paper, we develop a hybrid algorithm combined global optimization and gradient-based optimization methods to reconstruct optical parameters. Differential evolution is an effective global optimizer, and could yield an estimated solution near the global optimal region to eliminate the need of initial guess of the optical parameters. The gradient-based optimization method is further performed with initial values obtained from DE to obtain local optimal solution for more accurate reconstruction of optical parameters. This proposed reconstruction method is evaluated by comprehensive numerical simulation experiments. The results show that the proposed method is stable and accurate for the reconstruction of optical parameters.

## 1. Introduction

With the sensitive, specific, non-ionizing radiation and cost-effectiveness, the optical molecular imaging has widely applied for detecting molecular probes in small animals to monitor physiological processes, and visualizing cancer processes and therapeutic responses [1]. Fluorescent dyes or fluorescent reporter, and bioluminescence proteins are stable molecules, and are easily targeted to report on specific molecular processes in vivo. The planar fluorescence imaging is primarily qualitative and cannot resolve depth and quantify features. Optical microscopy techniques are applicable to study molecular and cellular processes in cell culture or tissue samples. Optical molecular tomographic imaging allows localizing and quantifying molecular probe distribution of in a living small animal using light intensity measurements on the animal body surface, and offers a superior ability to look deeper molecular and cellular activities in vivo inside biological tissues [2, 3].

The light intensity distribution on the tissue surface is related to the tissue anatomical structures, photon propagation model, and the optical properties of the tissues. The detailed anatomical structures of a small animal can be obtained by micro-MRI, x-ray micro-CT, or x-ray grating tomographic imaging. Photon propagation in biological tissue can be predicted by the radiative transfer equation (RTE), or Monte Carlo simulation. Optical parameters are also important for the optical molecular tomography. Distorted optical parameters would significantly affect the tomographic localization and quantifications of molecular probes [4]. However, the identification of optical properties is still a challenge due to its nonlinear and ill-posedness. Generally, this non-linear problem can be linearized to linear inverse problem using the Born approximation or the Rytov approximation, and is solved using an iterative method [5-7]. The optimization with regularization is a popular method for the optical tomography. Because there is multiple local optimal solutions in the optimization, the gradient-based optimization strongly rely on the initial choice, and could converge prematurely to an inaccurate solution if the initial guess is not close enough to the true solution. In addition, the linearized approximation errors and noisy measurement would result in an inaccuracy result of optical parameter reconstruction.

In this paper, a hybrid iterative method, combining differential evolution (DE) global optimization and gradient-based local optimization method, is proposed to reconstruct optical parameters of biological tissues. DE is an effective global optimization method operated on multiple sample points based on a spontaneous self-adaptive vector difference operator [8, 9]. This method displays a superior converging behavior, and can achieve a robust optical parameter reconstruction. However, DE needs to implement a large number of forward model computations, which take on more computational cost to enhance the accuracy of solution. To reduce the computational cost, DE is performed for less iteration to give an estimated solution near the global optimal region, and then the gradient-based optimization method is further performed with initial values obtained from DE method to obtain local optimal solution for more accurate reconstruction of optical parameters. The rest of the paper is organized as follows. In section 2, we provide a detailed description for the proposed reconstruction method of optical parameters. In section 3, we perform the numerical simulation experiments to demonstrate the hybrid algorithm for the reconstruction of optical parameters. Our numerical results validate the effectiveness of the proposed method. After that, we conclude the paper in the last section.

## 2. Methodology

**Geometrical model:** A small animal anatomy structures are complex, and can be reconstructed using structure imaging tools with high resolution, like micro-CT, micro-MRI, or x-ray grating tomographic imaging. One can segment the structural images into major organ regions, lungs, liver, heart, kidneys, stomach, bone and muscles, as shown in Fig. 1. The finite element method (FEM) is effective for representing an object with irregular external surface and complex internal structure. Because each organ region includes many subtle textures, it is difficult to utilize proper number of finite elements to present all the features in a finite element model. Otherwise, the computation for the forward/inverse photon propagation model would become impractical based on an overcomplicated geometrical model. A feasible way is to take the average optical parameter to describe the optical properties over each organ region. This technique assures that the boundary of every organ region is smooth to generate the qualified finite-element meshing, and facilitate the image reconstruction. Using the efficient 3D meshing techniques (Amira 4.0, Mercury Computer Systems, Inc. Chelmsford, MA), a high-quality finite-element mesh can be generated from segmented structural images to describe the geometrical shape and anatomic structure of a small animal.

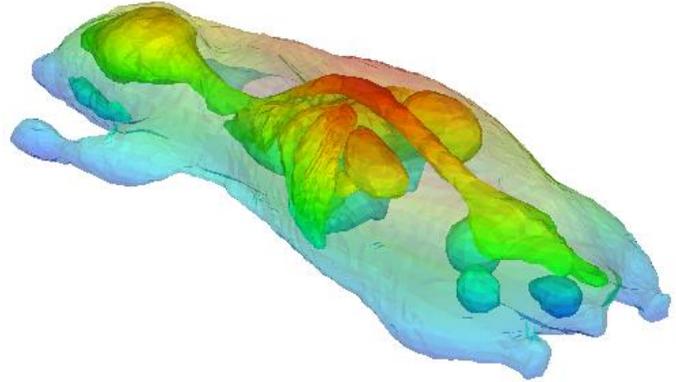

**Figure 1. Geometrical model of a mouse.**

**Diffusion approximation model:** Photon propagation in biological tissue is well described by the radiative transfer equation (RTE) [10]. The RTE is an integro-differential equation derived on the basis of the energy equilibrium in a scattering and attenuating medium. Monte Carlo simulation is a proven and popular stochastic method, which provides highly accurate solutions of the radiative transfer equation [11]. However, the high computational cost is a challenge for dealing with inverse problems in the biomedical imaging. For a small animal imaging, optical scattering is significantly dominant over the absorption in the biological tissues. The optical absorption coefficients of biological tissues is low in the near-infrared (NIR) window [12]. In this case, diffusion theory is of a high computational efficiency, and provides a quite accurate description for photon propagation to estimate photon fluence rates on the surface [13],

$$-\nabla \cdot (D(r)\nabla \Phi(r)) + \mu_a(r)\Phi(r) = 0, \quad r \in \Omega \tag{1}$$

where $\Omega$ is the region of interest in the imaging object, $\Phi(r)$ represents the photon fluence rates [Watts/mm$^2$] at location r; $S(r)$ the source density [Watts/mm$^3$]; $\mu_a(r)$ the absorption coefficient [mm$^{-1}$]; $D(r) = 1/[3(\mu_a + (1-g)\mu_s)]$ is the diffusion coefficient, which is related to the scattering coefficient $\mu_s(r)$ and the anisotropic scattering factor $g$. Based on the assumption that only an external light source $S(r)$ excites the object at boundary without any other background light, the boundary condition is derived by incorporating diffuse boundary reflection arising from a refractive index mismatch between $\Omega$ and the surrounding medium [13, 14],

$$(1-\alpha)\Phi(r) + 2(1+\alpha)D(r)\frac{\partial \Phi(r)}{\partial \nu} = S(r), \quad r \in \partial\Omega \tag{2}$$

where $\nu$ is an outward unit normal vector on $\partial\Omega$, and $\alpha$ the boundary mismatch factor between the tissue with a refractive index $n$ and air, which can be approximated by $\alpha = (1+\gamma)/(1-\gamma)$ with $\gamma = -1.44n^{-2} + 0.71n^{-1} + 0.67 + 0.06n$ [15].

**Solution of diffusion equation by finite element method:** The region of interested $\Omega$ is discretized into $p$ nonoverlapping tetrahedral elements $\Omega = \bigcup_{i=1}^{p} \Delta_i$, joined at vertex nodes $N_i$ $(i = 1, 2, \cdots, T)$. The photon fluence rates $\Phi(r)$ is approximated by the piecewise function $\Phi(r) = \sum_{i=1}^{T} \Phi_i \varphi_i(r) \in \Theta$, where $\Phi_i$ $(i = 1, 2, \cdots, T)$ is the photon fluence rate at node $N_i$. $\Theta$ is a finite-dimensional subspace spanned by basis function $\varphi_i(r), i = 1, 2, \cdots, T$, $\varphi_i(r)$ have support only over those elements $\Delta_k$ that have node $N_i$ as one of their vertices. The simplest basis is the piecewise linear basis, defined by $\varphi_i(N_j) = \delta_{ij}, i, j = 1, 2, \cdots, T$. This condition uniquely determines the basis function $\varphi_i(r)$ so that piecewise linear function that is 1 at $r = N_i$, and zero at all other nodes. Based on finite element analysis, a matrix equation in term of photon fluence rates can be established [16],

$$\mathbf{A}(D, \mu_a)\Phi = \mathbf{B}, \tag{3}$$

where

$$\begin{cases} \mathbf{A} = \begin{bmatrix} a_{11} & a_{12} & \cdots & a_{1T} \\ a_{21} & a_{22} & \cdots & a_{2T} \\ \vdots & \vdots & \cdots & \vdots \\ a_{T1} & a_{T2} & \cdots & a_{TT} \end{bmatrix}, \Phi = \begin{bmatrix} \Phi_1 \\ \Phi_2 \\ \vdots \\ \Phi_T \end{bmatrix}, \mathbf{B} = \begin{bmatrix} b_1 \\ b_2 \\ \vdots \\ b_T \end{bmatrix} \\ a_{i,j} = \int_\Omega \left( D(r) \nabla \varphi_i(r) \nabla \varphi_j(r) + \mu_a(r) \varphi_i(r) \varphi_j(r) \right) dr + \frac{1-\alpha}{2(1+\alpha)} \int_{\partial\Omega} \varphi_i(r) \varphi_j(r) dr \\ b_i = \frac{1}{2(1+\alpha)} \int_{\partial\Omega} \varphi_i(r) S(r) dr \end{cases} \quad (4)$$

The matrix $\mathbf{A}$ in Eq. (4) are positive definite. So we have $\Phi = [\mathbf{A}(D, \mu_a)]^{-1} \mathbf{B}$. Let $\Gamma$ denotes the boundary nodal set on the finite element mesh. The measurable photon fluence rates can be simulated by $\Phi(\Gamma, D, \mu_a)$, which is a function of scattering and absorption coefficients.

**Reconstruction of optical parameters:** The optical properties of a tissue are described in terms of the absorption coefficient and the scattering coefficient. Based on the finite element geometrical model described in above Section, the reconstruction of optical properties over organ regions can be performed from the measurements of photon fluence rates. In the optical imaging, near-infrared laser is used to irradiate the object (mouse) at different positions on object surface so that sufficient light signals transmit through major tissues. The transmitted diffusive light signal is measured on the opposite side of the mouse body. Hence, we can obtain excitation-diffusion measurement datasets using an optical imaging system. We further assume that the optical properties are constant in the same tissue type or organ region. As a result, this approach only deals with a much less number of unknown optical parameters than the traditional DOT approach does, which helps improve the numerical stability of the optical parameter reconstruction. In this case, the reconstruction of optical parameter is reduced to minimize the difference between model prediction and measurement data, which is expressed as the following optimization problem:

$$(D^*, \mu_a^*) = \mathrm{argmin} \sum_l \left\{ \| m_l - \Phi_l(\Gamma, D, \mu_a) \| \right\} \quad (5)$$

where $m_l$ is the measurement of photon fluence rates at the object surface and $\Phi_l(\Gamma, D, \mu_a)$ is corresponding photon fluence rates predicated by the diffusion approximation model. The objective function is a nonlinear and nonconvex. Generally, there are multiple local optimal solutions in this optimization. This optimization Eq. (5) can be solved using differential evolution (DE) method. DE is a global optimizer using population-based stochastic method. DE uses mutation as a search mechanism and selection to direct the search toward the prospective regions in the feasible region [8]. In DE, The solution vector with a dimension of $m$ consists of the optical parameters: $x = (D, \mu_a)$ [8, 9]. DE is a population based search technique which utilizes $NP = N \times G$ variables as population of parameter vectors for each generation. The collection of solution vectors is called the population denoted as,

$$P = \{x_{i,g}\}, \quad i = 1, 2, \cdots, N, \quad g = 1, 2, \cdots, G \quad (6)$$

where $N$ is the population size, and $G$ is the number of generations. Each of the $N$ solution vectors undergoes mutation, crossover and selection. The population is mutated using the unique differential operator to expand the search space:

$$v_{i,g+1} = x_{r_0,g} + f\left(x_{r_1,g} - x_{r_2,g}\right) \quad (7)$$

where $f$ is a differential weight factor, which is the most important control parameter of DE. The difference vectors $x_{r_1,g}$ and $x_{r_2,g}$ are randomly selected, and their weighted differential is used to perturb the base vector $x_{r_0,g}$. A crossover operation between the new generated mutant vector $v_{i,g+1}$ and a target vector $x_{t,g}$ selected sequentially from the population generate a trial vector $u_{i,g+1}$ to further increase the diversity of the new candidate solution:

$$u_{i,j,g+1} = \begin{cases} v_{i,j,g+1} & \text{if } rand(0,1) \leq C_r, \text{ or } j = Idx \\ x_{t,j,g} & \text{otherwise} \end{cases} \quad (8)$$

where $rand(0,1)$ is a randomly chosen real number in the interval $[0,1]$, $Idx$ is a randomly chosen integer number in the rang $[0,m]$, and $C_r$ is the crossover rate. The target vector $x_{t,g}$ is compared with the trial vector $v_{i,g+1}$ and the one with the lowest function value is selected to the next generation,

$$x_{i,g+1} = \begin{cases} u_{i,g+1} & \text{if } f(u_{i,g+1}) \leq f(x_{t,g}) \\ x_{t,g} & \text{otherwise} \end{cases} \quad i = 1, 2, \cdots, N \quad (9)$$

Finally, the above procedure is repeated for all *NP* parents to generate the next generation population. The solution vector with the smallest function value is used as the optimal solution when the maximum generation is reached.

To enhance the accuracy of solution, the solution $(D^*, \mu_a^*)$ generated by the DE optimization is used as the initial starting point for further gradient-based optimization. We represent true optical parameters as the sum between estimated optical parameters and differential parameters: $D = D^* + \delta D$ and $\mu_a = \mu_a^* + \delta \mu_a$. From the optical parameters $(D^*, \mu_a^*)$, we can calculate photon fluence rates $\Phi^*(r)$ based on diffusion approximation model:

$$\begin{cases} -\nabla \cdot (D^*(r)\nabla\Phi^*(r)) + \mu_a^*(r)\Phi^*(r) = 0, & r \in \Omega \\ (1-\alpha)\Phi^*(r) + 2(1+\alpha)D^*(r)\dfrac{\partial \Phi^*(r)}{\partial \nu} = S(r), & r \in \partial\Omega \end{cases} \quad (10)$$

For optical parameters $(D, \mu_a)$, the corresponding photon fluence rates $\Phi(r) = \Phi^*(r) + \delta\Phi(r)$ meets following diffusion approximation equation

$$\begin{cases} -\nabla \cdot ((D^*(r)+\delta D(r))\nabla(\Phi^*(r)+\delta\Phi(r))) + (\mu_a^*(r)+\delta\mu_a(r))(\Phi^*(r)+\delta\Phi(r)) = 0 \\ (1-\alpha)(\Phi^*(r)+\delta\Phi(r)) + 2(1+\alpha)(D^*(r)+\delta D(r))\dfrac{\partial(\Phi^*(r)+\delta\Phi(r))}{\partial \nu} = S(r), \quad r \in \partial\Omega \end{cases} \quad (11)$$

Combining Eq. (10-11) and ignoring higher order terms, we obtained,

$$\begin{cases} -\nabla \cdot (D^*(r)\nabla\delta\Phi(r)) + \mu_a^*(r)\delta\Phi(r) = \nabla \cdot (\delta D(r)\nabla\Phi^*(r)) - \delta\mu_a(r)\Phi^*(r) \\ (1-\alpha)\delta\Phi(r) + 2(1+\alpha)D^*(r)\dfrac{\partial(\delta\Phi(r))}{\partial \nu} = 0 \end{cases} \quad (12)$$

The governing equations (12) can be equivalently represented as the following weak form for any basis function $\varphi_i(r)$ ($i = 1, 2, \cdots, N$),

$$\int_\Omega D^*(r)\nabla\delta\Phi(r)\nabla\varphi_i(r) + \int_\Omega \mu_a^*(r)\delta\Phi(r)\varphi_i(r) + \frac{1-\alpha}{2(1+\alpha)}\int_{\partial\Omega} \delta\Phi(r)\varphi_i(r) = -\int_\Omega \delta D(r)\nabla\Phi^*(r)\nabla\varphi_i(r) - \int_\Omega \delta\mu(r)\Phi^*(r)\varphi_i(r)$$

(13)

According to finite element analysis, a matrix equation can be established from Eq. (13),

$$\mathbf{A}(D^*, \mu_a^*)\delta\Phi = -\mathbf{B}(\Phi^*)\delta D - \mathbf{C}(\Phi^*)\delta\mu_a, \quad (14)$$

where the components of the matrix $\mathbf{A}(D^*, \mu_a^*)$ is computed as

$$a_{i,j} = \int_\Omega D^*(\mathbf{r})\nabla\varphi_i(\mathbf{r})\cdot\nabla\varphi_j(\mathbf{r})d\mathbf{r} + \int_\Omega \mu_a^*(\mathbf{r})\varphi_i(\mathbf{r})\varphi_j(\mathbf{r})d\mathbf{r} + \frac{1-\alpha}{2(1+\alpha)}\int_{\partial\Omega}\varphi_i(\mathbf{r})\varphi_j(\mathbf{r})d\mathbf{r}, \quad i,j=1,2,\cdots,N, \qquad (15)$$

and the components of the matrix $\mathbf{B}$ and $\mathbf{C}$ is computed as

$$\begin{cases} b_{i,j} = \sum_{k=1}^{N}\Phi_k^*\int_\Omega \varphi_j(\mathbf{r})\nabla\varphi_k(\mathbf{r})\nabla\varphi_i(\mathbf{r})d\mathbf{r} \\ c_{i,j} = \sum_{k=1}^{N}\Phi_k^*\int_\Omega \varphi_i(\mathbf{r})\varphi_k(\mathbf{r})\varphi_j(\mathbf{r})d\mathbf{r} \end{cases} \quad i,j=1,2,\cdots,N \qquad (16)$$

Performing inverse operation of matrix $\mathbf{A}$ for the both sides of Eq. (13), we obtain,

$$\delta\Phi(\Gamma) = -\mathbf{A}^{-1}(\Gamma)\mathbf{B}(\Phi^*)\delta D - \mathbf{A}^{-1}(\Gamma)\mathbf{C}(\Phi^*)\delta\mu_a \qquad (17)$$

where $\delta\Phi(\Gamma) = m - \Phi^*(\Gamma)$, $\Gamma$ is a boundary nodal set on the finite element mesh corresponding to the measurable photon fluence rates. Eq. (14) is a linear system to describe the relationship between the measured photon fluence rate and optical parameters. Thus, based on the gradient-based optimization method, an iterative method can be performed from initial optical parameters $(D^*, \mu_a^*)$ to reach more accurate solution [17].

**Algorithm 1**

---

*Input data* $\{m_l : l=1,2,\cdots\}$

    *Establish a matrix equation using finite element method:* $\mathbf{A}(D,\mu_a)\Phi = \mathbf{B}$

    *Performing DE for global optimization:* $(D^*,\mu_a^*) = \text{argmin}\sum_l\{\|m_l - \Phi_l(\Gamma,D,\mu_a)\|\}$

*While* the current solution is not converged *do*

    *Establish a matrix equation using finite element method:* $\mathbf{A}(D^*,\mu_a^*)\delta\Phi = -\mathbf{B}(\Phi^*)\delta D - C(\Phi^*)\delta\mu_a$,

    *Performing local optimization:*

$$(\delta D^*, \delta\mu_a^*) = \text{argmin}\left[\|\delta\Phi(\Gamma) + \mathbf{A}^{-1}(\Gamma)\mathbf{B}(\Phi^*)\delta D + \mathbf{A}^{-1}(\Gamma)\mathbf{C}(\Phi^*)\delta\mu_a\|\right]$$

    *Update:* $D = D + \delta D^*; \quad \mu_a = \mu_a + \delta\mu_a^*$

*End While*

---

Table 1. Optical parameters reconstructed by DE

| Organ regions | Ground truth | | Reconstruction | |
|---|---|---|---|---|
| | $\mu_a\ (mm^{-1})$ | $\mu'_s\ (mm^{-1})$ | $\mu_a\ (mm^{-1})$ | $\mu'_s\ (mm^{-1})$ |
| muscle | 0.019 | 0.66 | 0.016 | 0.62 |
| spleen | 0.029 | 0.98 | 0.052 | 0.78 |
| lung | 0.068 | 0.95 | 0.058 | 0.87 |
| heart | 0.076 | 0.56 | 0.076 | 0.50 |
| liver | 0.047 | 0.58 | 0.039 | 0.47 |
| Stomach | 0.043 | 0.75 | 0.055 | 0.67 |

## 3. Numerical simulations

To test the accuracy of our algorithm, the simulation experiments are performed based on a digital mouse phantom, which includes 6 organ regions (muscle, lungs, heart, liver, stomach, and spleen) with appreciate optical parameters, as shown in Fig. 2. A finite element model was established from the micro CT image volume of a mouse [18]. The finite element model was composed of 5492 nodes and 26445 tetrahedrons with average edge length of 0.80mm. A laser with wavelength of 720nm was used as light source to excite down surface of the phantom, and optical detector was used to acquire photon fluence rate upper surface of the phantom, forming an excitation-detection dataset. This procedure is performed alternately for various excitation locations. Simulated photon fluence rate on boundary was generated using optical diffusion model. The Gaussian noise is added photon fluence rates to simulate real measurements.

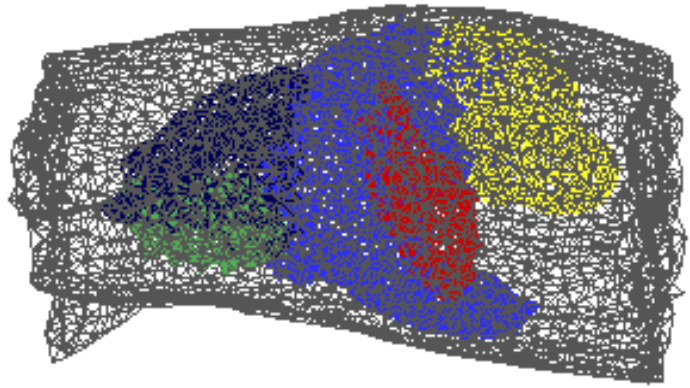

**Figure 2. Finite element model of a mouse phantom including 6 organ regions: muscle, lungs, heart, liver, stomach, and spleen.**

We used a population size of 60, differential scale factor of 0.8, and crossover rate of 0.5 for the differential evolution (DE) optimization and allow maximum 20 generations. With the setting of parameters, DE generates an optimal solution, which was listed in Table 1. To further enhance the accuracy of solution, the quasi-Newton method was performed with the initial values generate by the DE optimization. The reconstruction results are listed in Table 2. The relative error for the reconstruction of $\mu_a$ is less than 30% and the relative error for $\mu_s'$ is less than 13%. The differential evolution is robust to noise data and does not request the initial guess of the optical parameters. The proposed optical property characterization method improves the accuracy for optical molecular imaging modalities. It should be pointed out that the accuracy of optical parameters significantly rely on the volumetric size of organ regions, the bigger volumetric size of organ regions, the more accurate the reconstructed optical parameter will be. Because smaller organ regions would has less contribution to the boundary measurements of transmittance intensity, the deviation of optical parameters would not generate considerable effect to optical molecular tomography.

Table 2. Optical parameters reconstructed by gradient-based optimization method

| Organ regions | Ground truth | | Reconstruction | |
|---|---|---|---|---|
| | $\mu_a \left(mm^{-1}\right)$ | $\mu_s' \left(mm^{-1}\right)$ | $\mu_a \left(mm^{-1}\right)$ | $\mu_s' \left(mm^{-1}\right)$ |
| muscle | 0.019 | 0.66 | 0.017 | 0.67 |
| spleen | 0.029 | 0.98 | 0.044 | 0.91 |
| lung | 0.068 | 0.95 | 0.069 | 0.95 |
| heart | 0.076 | 0.56 | 0.079 | 0.54 |
| liver | 0.047 | 0.58 | 0.058 | 0.54 |
| Stomach | 0.043 | 0.75 | 0.052 | 0.85 |

4. **Conclusion and discussions**

In conclusion, we have presented hybrid iterative method, combining differential evolution global optimization and gradient-based local optimization method, to quantify the optical parameters of biological tissue based on the diffusion approximation model. Numerical simulation experiments have demonstrated that this reconstruction method is accuracy, stable and robust against measured noise data. The proposed reconstruction method of optical parameters would improve the accuracy for optical molecular tomographic imaging. Furthermore, small animal experiments will be carried out to combine the proposed method with optical molecular tomography for preclinical imaging studies.

The reconstruction of optical parameters is rather complex which is related to many kind of factors, for example the laser excitation modes, organ tissue segmentation of structure images, photon propagation model, and the signatures of measurement data. The structural image of the small animal can be reconstructed using computed tomography (CT) or magnetic resonance imaging (MRI). With the prior knowledge of the structural information, reconstruction of optical parameters is significantly simplified for more efficient and robust algorithm design. While there is not an analytical method available for the reconstruction of optical parameters, optimization method is feasible option. However, the objective function is not convex, and there are multiple local optimal solutions in this optimization. The proposed hybrid iterative approach eliminates the need of initial guess of the optical parameters in conventional construction methods, and would not take large computation burden as well. This proposed reconstruction method is based on the diffuse approximation model for high computational efficiency. It is also straightforward to extend our method to more accurate photon transport models, such as radiative transport equation (RTE) [19], phase approximation [20], or Monte Carlo simulation [2]. Although the iterative reconstruction is time-consuming, several forward solutions can be calculated at the same time, and the running time for the iterative process is reduced to a reasonable time frame with the aid of parallel programming on a multi-core workstation. The iterative reconstruction has the advantage to allow much freedom in the data acquisition design. The illumination source not only can be placed on the boundary of an object, but inside the object as well. The photon escaped from the surface of the object can be captured either directly by a CCD camera or through an optical fiber.

## Acknowledgment

This work was supported by the National Institutes of Health Grant NIH/NIBIB R01 EB019443